\documentclass[12pt,preprint]{aastex}

\slugcomment{
A. Asai : 2012.11.15
}

\shorttitle{Spectral Indices for Nonthermal Emissions}
\shortauthors{Asai et al.}

\begin{document}

\title{Temporal and Spatial Analyses of Spectral Indices of Nonthermal
Emissions Derived from Hard X-Rays and Microwaves}

\author{
Ayumi Asai\altaffilmark{1}, 
Junko Kiyohara\altaffilmark{2}, 
Hiroyuki Takasaki\altaffilmark{2,3}, 
Noriyuki Narukage\altaffilmark{4}, 
Takaaki Yokoyama\altaffilmark{5}, 
Satoshi Masuda\altaffilmark{6}, 
Masumi Shimojo\altaffilmark{7}, and 
Hiroshi Nakajima\altaffilmark{7}}

\email{asai@kwasan.kyoto-u.ac.jp}
\altaffiltext{1}{
Unit of Synergetic Studies for Space, Kyoto University,
Yamashina, Kyoto, 607-8471, JAPAN}

\altaffiltext{2}{
Kwasan and Hida Observatories, Kyoto University,
Yamashina, Kyoto, 607-8471, JAPAN}

\altaffiltext{3}{
acwest, Inc., Shibuya, Shibuya, Tokyo, 150-0002, JAPAN}

\altaffiltext{4}{
Institute of Space and Astronomical Science, Japan Aerospace Exploration
Agency, Chuo, Sagamihara, Kanagawa, 229-8510, JAPAN}

\altaffiltext{5}{
Department of Earth and Planetary Science, University of Tokyo, 
Hongo, Bunkyo, Tokyo, 113-0033, JAPAN}

\altaffiltext{6}{
Solar-Terrestrial Environment Laboratory, Nagoya University, Chikusa, 
Nagoya, Aichi, 464-8601, JAPAN}

\altaffiltext{7}{
Nobeyama Solar Radio Observatory, National Astronomical Observatory of
Japan, Minamimaki, Minamisaku, Nagano, 384-1305, JAPAN}

\begin{abstract}
We studied electron spectral indices of nonthermal emissions seen in
hard X-rays (HXRs) and in microwaves.
We analyzed 12 flares observed by the Hard X-ray Telescope aboard {\it
Yohkoh}, Nobeyama Radio Polarimeters (NoRP), and the Nobeyama
Radioheliograph (NoRH), and compared the spectral indices derived from
total fluxes of hard X-rays and microwaves.
%%%
Except for four events, which have very soft HXR spectra suffering from
the thermal component, these flares show a gap $\Delta\delta$ between
the electron spectral indices derived from hard X-rays $\delta_{X}$ and 
those from microwaves $\delta_{\mu}$ ($\Delta\delta = \delta_{X} -
\delta_{\mu}$) of about 1.6.
Furthermore, from the start to the peak times of the HXR bursts, the
time profiles of the HXR spectral index $\delta_{X}$ evolve
synchronously with those of the microwave spectral index $\delta_{\mu}$,
keeping the constant gap.
%%%
We also examined the spatially resolved distribution of the microwave
spectral index by using NoRH data.
The microwave spectral index $\delta_{\mu}$ tends to be larger, which
means a softer spectrum, at HXR footpoint sources with stronger
magnetic field than that at the loop tops.
These results suggest that the electron spectra are bent at around
several hundreds of keV, and become harder at the higher energy range
that contributes the microwave gyrosynchrotron emission.
\end{abstract}

\keywords{Sun: flares --- Sun: corona --- Sun: radio radiation --- Sun:
X-rays, gamma rays --- acceleration of particles}

\section{Introduction}
%%%
During the impulsive phase of solar flares, electrons are accelerated up
to the energy of MeV, and are responsible for nonthermal emissions with
power-law spectra observed in hard X-rays (HXRs) and in microwaves.
%The nonthermal electrons often show a power-law distribution.
In HXRs the bremsstrahlung emission from electrons with the energy of
more than tens of keV is dominant \citep{Brown71}, while the
gyrosynchrotron emission from more than several hundreds of keV
electrons is dominant in microwaves \citep{Whi92,Bas98,Bas99}.
HXR and microwave nonthermal emissions are very similar in the light
curves (e.g., Kundu 1961; Kai 1986), and therefore, it has been
generally accepted that they are produced by a common population of
accelerated electrons.

The electron spectra derived from HXR emissions are, on the other hand,
often different from those derived from microwave emissions.
The electron spectral indices of the power-law distribution derived from
HXRs $\delta_{X}$ are, in many cases, larger than those derived from
microwaves $\delta_{\mu}$, that is $\delta_{X} > \delta_{\mu}$
\citep{Kun65,Sil97,Sil00}.
According to the study by \citet{Sil97}, the average gap between
$\delta_{X}$ and $\delta_{\mu}$, $\Delta \delta$ ($= \delta_{X} -
\delta_{\mu}$) is 0.5 -- 2.0.
%%%
This means that microwave spectra are harder than the HXR ones, and that
acceleration and/or traveling mechanisms could be different for these
wavelengths.
For example, \citet{Mel76} suggested the so-called
trap-plus-precipitation model, in which the magnetic trap works more
effectively for such higher energy electrons that emit microwaves than
HXR emitting electrons with lower energy.
\citet{Mino08}, on the other hand, showed in their numerical calculation
that the gap $\Delta\delta$ is naturally generated even from the
electron distribution with a single power-law spectrum, since microwave
and HXR emissions are from different electrons (trapped and
precipitating ones, respectively.)
%%%
In the early phase of a flare, however, the magnetic trap is probably
not so effective, and we can examine the features of the nonthermal
electrons without having to deal with the trapping effect.
Moreover, it is crucially important to analyze imaging data both in HXRs
and in microwaves, since we can resolve spatially the precipitating (at
footpoints) and trapped (at loop tops) components with them.

In this paper, we report the results of the analyses on the electron
spectral indices derived from HXRs and microwaves, and also discuss the
temporal and spatial characteristics.
We used the HXR data obtained with the Hard X-ray Telescope (HXT; Kosugi
et al. 1991) aboard the {\it Yohkoh} satellite \citep{Oga91}.
The microwave data were taken with the Nobeyama Radio Polarimeters
(NoRP; Torii et al. 1979; Shibasaki et al. 1979; Nakajima et al. 1985a),
and the Nobeyama Radioheliograph (NoRH; Nakajima et al. 1994) at
Nobeyama Solar Radio Observatory, National Astronomical Observatory of
Japan.
These data enable us to examine the nonthermal features of the
accelerated electrons spatially, temporally, and spectroscopically both
in HXRs and in microwaves.
%%%
In \S2 and \S3 we summarize the observations and the method of the
analyses, respectively, and we show the results of the statistical
analyses in \S4.
In \S5 we present discussions and our conclusions.

\section{Observations}
%%%
We used HXR data obtained with {\it Yohkoh}/HXT.
By using the HXT data, we can synthesize the HXR images in four energy
bands, namely the L band (14 -- 23~keV), M1 band (23 -- 33~keV), M2 band
(33 -- 53~keV), and H band (53 -- 93~keV).
The spatial and temporal resolutions of the HXT images are
5$^{\prime\prime}$ and 0.5~s, respectively.
%%%
To obtain the HXR photon spectral index $\gamma$ ($I_{X}(\epsilon)
\propto \epsilon^{-\gamma}$, where $I_{X}$ is the HXR intensity, and
$\epsilon$ is energy of the photon), we used the data in the two highest
energy bands, i.e., the HXT M2 and H bands, because HXR emissions with
energy less than 30~keV sometimes suffer from the contribution of the
thermal emissions.
Therefore, the HXR photon spectral index $\gamma$ is roughly written as
$- \log(I_{X}({\rm H})/I_{X}({\rm M2})) (\log(\epsilon_{\rm
H}/\epsilon_{\rm M2}))^{-1}$, where $\epsilon_{\rm M2}$ and
$\epsilon_{\rm H}$ are the effective energies for the M2 and H bands.
Though, as we will discuss later, some events showed the effects of the
super-hot thermal components even in the M2 band.
In this work we calculated the HXR photon spectral indices by using the
programs \verb"hxt_powerlaw" in the Solar SoftWare (SSW) package on IDL.
We accumulated the HXT data for two seconds in the spectral analyses to
reduce the photon noise.

We also used the HXR spectra derived from the Wide Band Spectrometer
(WBS; Yoshimori et al. 1991).
The Hard X-ray Spectrometer (HXS), one of the sensors installed on the
WBS, is dedicated to taking HXR spectra over a wide energy range.
WBS/HXS can obtain HXR spectra with higher spectral resolution than HXT,
although it cannot obtain the spatial information.
Therefore, the data of WBS/HXS and HXT are complementary for the imaging
spectroscopy in HXRs.
\citet{Sato06} summarized all the events observed by WBS, and we used
their database of the HXR spectra.

In microwaves the gyrosynchrotron emission is dominant during the
impulsive phase of a flare.
The spectrum in microwave range $F_{\nu}$ is approximately fitted with
the two power-law indices $\alpha_{\rm tk}$ and $\alpha_{\rm tn}$ by the
function as follows:
\begin{equation}
F_{\nu} = F_{\nu, pk}
\left(\frac{\nu}{\nu_{pk}}\right)^{\alpha_{\rm tk}}
\left\{
1-\exp\left[
-\left(\frac{\nu}{\nu_{pk}}\right)^{\alpha_{\rm tn}-\alpha_{\rm tk}}
\right]
\right\}
\approx
\left\{
\begin{array}{cc}
F_{\nu, pk}
\left({\nu}/{\nu_{pk}}\right)^{\alpha_{\rm tk}} 
  & \mbox{for $\nu \ll \nu_{pk}$}\\
F_{\nu, pk}
\left({\nu}/{\nu_{pk}}\right)^{\alpha_{\rm tn}} 
  & \mbox{for $\nu \gg \nu_{pk}$}
\end{array}
\right.
\label{equation:fitting}
\end{equation}
The optically thin part, therefore, follows the power law distribution
with a negative spectral index $\alpha$ ($= \alpha_{\rm tn}$) as
$F_{\nu} \propto \nu^{\alpha}$, where $F_{\nu}$ is the flux density at
frequency $\nu$ (e.g. \cite{Rama69,Dulk85}).

In this study we used the microwave data taken with NoRP, which measures
the total fluxes at 1, 2, 3.75, 9.4, 17, 35, and 80~GHz, with a temporal
resolution of 0.1~s.
By fitting a NoRP spectrum with the equation (1), we can obtain the
spectral index $\alpha_{P}$ for the optically thin gyrosynchrotron.
%%%
We did not use the NoRP 80~GHz data in this work due to the poor
statistics.
For some events, the NoRP 1~GHz data were also ruled out, because they
clearly did not follow the gyrosynchrotron, but instead, the plasma
emission.
We accumulated the NoRP data for 5~s to reduce the noise level.
The microwave fluxes of the NoRP data are determined with the error of
less than 10~\% of the signal for 1, 2, 4, 9.4 GHz, and 15~\% for 17 and
35 GHz, which is mainly due to the observation conditions such as the
calibration and the weather.
The error to determine $\alpha_{P}$ is also affected by the accuracy of
fitting, and is finally estimated to be about $\pm$ 0.5.

NoRH also observes the sun at 17 and 34~GHz.
The microwave 2-dimensional images are synthesized from the NoRH data,
and the spatial resolutions (FWHMs of the synthesized beam) of about
14$^{\prime\prime}$ for 17 GHz and 7$^{\prime\prime}$ for 34~GHz.
The time cadence of the data we used in this work is one second.
%%%
We can derive the two-dimensional distribution (map) of the microwave
spectral indices using the NoRH data ($\alpha_{H}$), by calculating
$\log(F_{34 \mathrm{GHz}}/F_{17 \mathrm{GHz}}) (\log(34 \mathrm{GHz}/17
\mathrm{GHz}))^{-1}$ for each position of images.
If we assume that the turnover frequency is less than 17~GHz, the derive 
$\alpha_{H}$ is for the optically thin gyrosynchrotron emission.

Here, we have to note the calibration of the NoRH 17 and 34~GHz data.
NoRH calibrates phase and gain by using the sun itself (i.e. the solar
disk) as a calibrator, thanks to the redundant antenna configuration.
However, the fundamental (smallest) spacing of the antennas ($=1.5$~m)
corresponds to the maximum wavelength in the space of 20$^{\prime}$ ($=
1200^{\prime\prime}$) at 34~GHz, which means that the whole solar disk
is not resolved.
The solar disk at 34~GHz is partially overlapped with other fake solar
disk images, and therefore, the background (quiet) solar disk is
possibly not well determined.
%%%
To correct this, we adjusted the flux of the flare region taken by NoRH
34~GHz, which is subtracted by the preflare data, to the fitting
results derived from NoRP.
NoRP calibrates those fluxes by using sky and absorber levels.
In our cases the NoRH 34~GHz fluxes are usually smaller than the NoRP
35~GHz ones, and the ratio (NoRH 34~GHz)/(NoRP 35~GHz) is from 0.4 to
1.2.
This calibration possibly reduces the derived spectral index
$\alpha_{H}$ about $-$1.0 at a maximum.
We also corrected the NoRH 17~GHz flux of the flare region, by using the
fitting result from NoRP.
Although the correction ratio (NoRH 17~GHz)/(NoRP 17~GHz) ranges from
0.6 to 1.2, it is roughly comparable to 1 in most cases.
This calibration causes the error on $\alpha_{H}$, mainly due to the
measurement error of the radio flux by NoRP, and is about 0.4.
%%%%
The relative displacement between the images in 17~GHz and those in the
34~GHz due to the NoRH image syntheses also causes the error to estimate
the spectral index $\alpha_{H}$.
The NoRH image syntheses hold an uncertainty on the positioning of about
$5^{\prime\prime}$, and in this case, the error on $\alpha_{H}$ is about
$\pm$ 0.2 for bright emission sources.

\section{Data Analyses}
%%%
Strong (i.e., intense) and large events are preferable for our imaging
spectroscopic analyses.
Therefore, we sought solar flares observed by HXT, NoRH, and NoRP
for the period from the start of the dual-frequency observation with
NoRH (November 1995) to the end of the observation of {\it Yohkoh}
(December 2001), and selected 12 flare events that meet the following 
criteria; 
(1) The flare is larger than M1.0 on the GOES scale.
(2) The flare is an event listed in {\it The Yohkoh HXT/SXT Flare
Catalogue} \citep{Sato03}, and the maximum HXR intensity is so strong
that the counts per second per subcollimator in the HXT M2 band are
larger than 30.
(3) The spatial size of the microwave emission source observed with NoRH
at 17~GHz is large enough, and it is more than 4 times of the beam size.
(4) The microwave images can be successfully synthesized from the NoRH
data, in other words, the microwave emissions are not extremely
strong\footnote{The normal tool for synthesizing NoRH images often fails
in the case of very strong microwave emission.}.
(5) The optically thin parts of the gyrosynchrotron emission are well
defined from the NoRP data, which means that the turn-over frequencies
determined from the fitting do not reach 17 GHz or more.
(6) The peak time of the event is between 00:00~UT to 06:00~UT, which
corresponds to 09:00 and 15:00~JST (i.e. Japanese daytime), and the
NoRH beam pattern is not so distorted.
%%%
Table~1 lists all the 12 flares.
Figure~1 also shows temporal and spatial features of each event.
The selected events are widely distributed longitudinally and in the
flare size.

\subsection{HXR Spectral Indices}
%%%
In the selected events we confirmed that the dominant HXR emissions are
from footpoints of flare loops, and no loop top HXR sources are
included, such as those reported by \citet{Masu94}.
This means that they are produced by bremsstrahlung caused by the
interactions between the precipitating energetic electrons and the dense
chromospheric plasma.
Therefore, it is reasonable that we adopt the thick-target model
\citep{Brown71,Hudson78,Sakao94} for the HXR emission sources.
In this paper we determined the spectral indices $\gamma$ from the total
HXR intensities, assuming that the HXR emissions from the footpoint
sources are so strong that we can approximately equate the total
intensities to the emissions from the footpoint source.

The thick-target model suggests the relation between $\gamma$ and the
spectral index $\delta'_{X}$ of the number flux of the injected
energetic electrons $F_{N}(E)$ ($= dN(E)/dt \propto E^{-\delta'_{X}}$),
as ${\delta'_{X}}={\gamma} + 1.0$.
To estimate the spectral index $\delta_{X}$ of the accelerated electron
$N(E)$, we further have to consider the traveling time of the electrons
$\tau$ ($N(E) \approx F_{N}(E) \tau \propto E^{-\delta_{X}}$), since
energy-dependent $\tau$ moderates the spectral index and $\delta'_{X}
\neq \delta_{X}$.
%%%
In this paper we adopted the typical timescale over which the
precipitating electrons travel the effective length $L$ in the flare
loops with the velocity $v$, that is, $\tau = L v^{-1}$.
The traveling time is proportional to $v^{-1}$, that is, $\tau \propto
E^{-0.5}$.
Then, the electron spectral index derived from the HXR emissions
$\delta_{X}$ follows the relation, $\delta_{X} = \gamma + 1.5$.

The seventh column of Table~1 shows the derived spectral index
$\delta_{X}$ at a time when the HXR emission records the maximum in the
HXT M2 band for each event as shown in the second column.
The derived $\delta_{X}$ ranges from 3.8 to 6.6.
%%%
The error to estimate the spectral index $\gamma$ is about $\pm$ 0.5, and
the same is true for $\delta_{X}$, under the assumption of single
power-law HXR spectra.
The photon noise is the main factor of the error, and therefore, it is
even reduced during strong HXR emissions (to about $\pm$ 0.2).

As shown in Figure~1, in most cases, the temporal variations of the
spectral index, which are plotted with the cross mark ($+$), show the
soft-hard-soft (SHS) behavior at the peak times of the HXR emissions.
Therefore, $\delta_{X}$ recorded in Table~1 is roughly the minimum
value during the burst.
%%%
As we describe below, for the 2000 January 12 (event b) and the 2000
March 3 (event f) flares, we chose other sub-peaks instead of the
maximal intensities, because we could not well fit the microwave
spectra.

In Figure~2 we show the HXR spectra derived from WBS/HXS.
These are from the database of \citet{Sato06}.
The times on the top are the integration time in UT, and they almost
cover the HXR maxima.
The solid lines are the power-law distribution of the HXR spectra
derived from HXT, while the absolute values are arbitrary.
For the 2000 March 3 (event f) and the 2001 March 30 (event k) flare,
the integration times of WBS/HXS are out of the HXT peak times we took.
We can confirm that the HXR spectral indices derived from HXT fit the
HXR spectra of WBS/HXS within the margin of errors as small as $\pm$ 0.2 \citep{Sato06}.

\subsection{Microwave Spectral Indices}
%%%
Next, we derived the spectral indices of the accelerated electrons
$\delta_{\mu}$ from the microwave emissions taken with NoRP.
%%%
For the optically thin gyrosynchrotron emission, $\alpha$ is related to
the spectral index of the accelerated electrons $\delta_{\mu}$.
There have been several studies to derive the relation between
$\delta_{\mu}$ and $\alpha$, and we adopt the approximation derived by
\citet{Dulk85} here, and  $\delta_{\mu} = (1.22 - \alpha) / 0.9$.
In this paper, we distinguish the spectral index derived from NoRP
$\delta_{\mu_{P}}$ from that from NoRH $\delta_{\mu_{H}}$ with the
subscripts {\it P} and {\it H}.
%%%
Figure~3 shows some examples of the NoRP spectra and the fitting result at 
the time of HXR maximal intensity (column 2).
The eighth column of Table~1 also shows the spectral index derived from
NoRP $\delta_{\mu_{P}}$ at the time of the HXR maxim.
As we mentioned above (\S 2), the error to determine $\alpha_{P}$ is
about $\pm$~0.5, and therefore, that for $\delta_{\mu_{P}}$ is about
$\pm$~0.6.
%%%
We confirmed that the contribution from the free-free emission at the
35~GHz is negligible.
%%%
The temporal variations of the $\delta_{\mu_{P}}$ follow the
gradual-hardening (GH) or soft-hard-harder (SHH) behaviors as shown in
Figure~1.
%%%
We added the gap between the spectral index derived from HXRs and that
from microwaves $\Delta\delta = \delta_{X} - \delta_{\mu_{P}}$ in the
ninth column.
We will discuss $\Delta\delta$ in more detail later.

For the 2000 January 12 and 2000 March 3 flares (events b and f,
respectively), the turnover frequencies are higher than 20~GHz at the
HXT maxima, and we could not well fit the spectral indices $\alpha$
for the optically thin part of the gyrosynchrotron emission.
Therefore, we chose other sub-peaks of the HXR light curves of these
events for the spectral analysis.
These peak times are shown with gray vertical lines in Figure~1.

From the NoRH data, we can derive the spatially resolved spectral
indices $\alpha_{H}$ (and therefore, $\delta_{\mu_{H}}$).
We can compare, for example, $\delta_{\mu_{H}}$ at the footpoint sources
with that of the loop-top ones.
%%%
As we just mentioned, we confirmed that the turnover frequencies derived
from NoRP are lower enough than 17~GHz except for the peak times of the
events (b) and (f).
For the two events, we again selected preceding sub-peaks for further
spectral analyses.
Therefore, the indices $\alpha_{H}$ derived from only two frequencies
(17 and 34 GHz) are mainly for the optically thin gyrosynchrotron, while
we cannot rule out the possibility that they are locally enhanced to be
close to or larger than 17~GHz.
%%%
Figure~1 shows such two-dimensional distribution of the
index-$\alpha_{H}$, which we call ``$\alpha$-map'' in the bottom right
sub-figure.

\section{Results}
%%%
Table 1 lists the selected 12 events, with the peak time of the HXT M2
band (column 2), the delay of the peak times of microwave (NoRP 17~GHz)
compared to those of HXR (HXT M2 band) in second (column 3), the
spectral index derived from HXT $\delta_{X}$ (column 7), that derived
from NoRP $\delta_{\mu_{P}}$ (column 8), and the gap $\Delta\delta$ ($=
\delta_{X} - \delta_{\mu_{P}}$) (column 9).
%%%
In Figure~1 we present the time profiles of the HXR flux of HXT M2 band
and the microwave flux at 17~GHz taken with NoRP (top left).
The time profiles of the spectral indices $\delta_{X}$ and
$\delta_{\mu_{P}}$ are also shown with the cross ($+$) and square
($\square$) signs, respectively, in the bottom left sub-figure.
We also present the contour images in the microwave taken with NoRH at
17~GHz (top right) of the selected events.
The contour levels of the microwave images are 20~\%, 40~\%, 60~\%,
80~\%, and 95\% of the peak intensity.
The microwave images are overlaid with the HXR contour images taken with
HXT M2 band with the contour levels of 20~\%, 40~\%, 60~\%, 80~\%, and
95\% of the peak intensities.

\subsection{Peak Delay of Microwave Emissions}
%%%
As the top left sub-figures of Figure~1 show, the microwave peaks almost
always delay from the HXR peaks.
The delay times for the events are listed in the third column of
Table~1.
These delays of microwave peaks to HXR ones have been often observed,
and reported by various authors (e.g., G\"{u}del et al. 1991).
In the current case the delays are from 0 to 21 seconds, and the average
is 7.3 seconds.
These are comparable with the result by \citet{Naka85b}.
Delays of microwave peaks to HXT ones are thought to be caused by
magnetically trapping for microwave-emitting electrons, while HXR
emissions are from the directly precipitated electrons to the
chromosphere.

\subsection{Gap of Spectral Indices $\Delta\delta$}
%%%
According to the results summarized in Figure~1 and Table~1, especially
concerning the gap $\Delta\delta$, we first categorize these events into
two groups.
%%%
As we mentioned above, the temporal variations of $\delta_{X}$ clearly
show the SHS behavior at the peaks, while those of $\delta_{\mu_{P}}$
show hardening features (GH or SHH) as flares progress.
In most cases, therefore, the listed gap $\Delta\delta$ is the smallest
value near the HXR peaks.
For all events, the gap $\Delta\delta$ is always positive, which means
that the electron spectra derived from microwaves of this group are
always harder than that from HXRs.

The first is the group with smaller gap ($\Delta\delta <$ 2.2).
This group shows the electron spectral index derived from HXRs
$\delta_{X}$ of about 4.6 $\pm$ 0.8, while that derived from microwaves
$\delta_{\mu_{P}}$ is about 3.0 $\pm$ 0.8.
The gap $\Delta\delta$ of about 1.6 is well consistent with the result
by \citet{Sil00}.
The eight events (from a to h) belong to this group.
%%%
The other 4 events (from i to l), which belong to the second group, have
much larger gaps $\Delta\delta$ of greater than 2.7.
%%%
In particular, the electron spectral index derived from HXRs
$\delta_{X}$ is much larger than those of the first group, and is about
6.2 $\pm$ 0.5.
We checked the HXT data for those events, and concluded that the HXR
emissions suffered from the thermal components even in the M2 band.
The HXR spectra derived from WBS/HXS (Fig.~2) are also very soft.
%%%
\citet{Huang06} studied one of them (the 1998 November 28 flare; event
i), in more detail, and concluded that there is a vast super-hot thermal
component in this flare, which softens the HXR spectrum.
Henceforth, we discuss only the first group.

Even after we eliminate the second group, there remains the gap
$\Delta\delta$.
%%%
Figure~4 shows the scatter plot with the horizontal axis of $\delta_X$
and the vertical axis of $\delta_{\mu_{P}}$.
The events are marked with the asterisks ($\ast$).
The solid line shows the points where $\delta_X$ is equal to
$\delta_{\mu_{P}}$.
This also suggests that there is a certain gap between the spectral
indices $\Delta\delta$.

\subsection{Microwave Emission Sources}
%%%
As shown in Figure~1, the sites of the dominant microwave emissions
are different for each event.
%%%
According to the spatial displacement between the brightest position of
the microwave emission sources and those of the HXR sources, and/or
by using imaging observations in soft X-rays and in extreme ultraviolets,
we categorized the microwave emission sources into two cases; footpoint
source (group F; displacement is less than 5$^{\prime\prime}$) and loop
source (group L; displacement is larger than 5$^{\prime\prime}$) as
noted in the sixth column of Table~1.
Here, we note that the brightest microwave emission source seen in the
2000 November 25 flare (event g) is confirmed to be a footpoint source
\citep{Taka07}, although it is displaced from the HXR source of about
40$^{\prime\prime}$.
Therefore, we categorized this case as a footpoint event.

We confirmed that the dominant microwave emission sources appear at the
tops or legs of flare loops for many cases (5 of 8 events).
This is consistent with the result of \citet{Huang09a} or \citet{Mel02}.
%%%
We will discuss the difference between the temporal evolution of the
spectral indices for group L and that for group F in the next
subsection.

\subsection{Temporal Evolution of Spectral Indices}
%%%
We examine the temporal evolution of the spectral indices.
The HXR spectral index $\delta_{X}$ mostly shows the SHS behaviors at
each short peak, while two of them (events f and g) show even GH
features.
The temporal evolution of the microwave spectral index
$\delta_{\mu_{P}}$, on the other hand, seems different, according to the
position of microwave emission sources (i.e. group L/F).

For group L (events a, b, c, d, and e), the time profiles of the
spectral indices ($\delta_{\mu_{P}}$ and $\delta_{X}$) show similar
evolution from start to peak of the bursts, while the gap $\Delta\delta$
increases with time after the peaks.
In other words, the electron spectral index derived from HXRs
$\delta_{X}$ increases after the peaks, showing the SHS features, while
the electron spectral index from microwaves $\delta_{\mu_{P}}$ becomes
further smaller showing SHH or GH features.
%%%
For group F (events f, g, and h), on the other hand, both the
spectral indices ($\delta_{\mu_{P}}$ and $\delta_{X}$) show similar
evolution through the impulsive phase while maintaining a gap of about
1.6.
However, event (h) is different from the other two events of group F.
This event shows the SHS features both in the microwave index
$\delta_{\mu_{P}}$ and in the HXR index $\delta_{X}$, while the others
show GH features.

From the point of the temporal evolution, we can, therefore, re-classify
the eight events into the three groups: (L) microwave emission sources
are located at loop tops/legs, and $\delta_{X}$ and $\delta_{\mu_{P}}$
show SHS and GH features, respectively (events a, b, c, d, and e), (F)
microwave emission sources are from footpoints, and both $\delta_{X}$
and $\delta_{\mu_{P}}$ show GH features (events f and g), (F$^{*}$)
microwave emission source is again from footpoints, but both
$\delta_{X}$ and $\delta_{\mu_{P}}$ show SHS features (event h).

Figure~5 is roughly a realignment of Figure~1, but much more focuses on
the temporal evolution of the microwave spectral index
$\delta_{\mu_{P}}$ during the bursts for the groups (L; top) and (F;
bottom).
The vertical dotted lines correspond to the times when the microwave
17~GHz fluxes taken by NoRP exceeds 100~SFU (Solar Flux Unit).
We determine this time as the start of the bursts.
The vertical dashed lines correspond to the times when the 17~GHz fluxes
become half of the peak values.
We define this time as the end of the bursts.
Although even after the end times the microwave indices
$\delta_{\mu_{P}}$ become smaller, the fluxes probably suffer from the
thermal component in the microwave emissions, which should be avoided
from the analysis.
For the 2000 November 25 flare (event g), we defined the end time just
after the peak HXR time, since the hardening seems saturated after
01:12~UT, and no further hardening is seen.
%%%
The bottom panels again show the time profiles of the spectral index
$\delta_{\mu_{P}}$ with the square ($\square$) marks.
The overlaid time profiles are the spectral index $\delta_{X}$ (cross
$+$).
We subtracted 1.5 from the original value of the index to clearly show
the temporal variation.
%%%
The numbers noted in the figure show how much the microwave spectral
indices $\delta_{\mu_{P}}$ decrease during the two vertical dotted
lines.
We also summarized the hardening degree in the tenth column of Table~1.
As a result, the nonthermal spectra harden with the indices of 1.0 on
the average.
%%%
The hardening with the microwave spectral index of about 1 seems to be
consistent with the discussion of the trapping time of relativistic
electrons \citep{Bai79,Petro85}.
%it is described by the life time that depends on the energy ($t \propto
%E$), not by the deflection time \citep{Bai79,Petro85}.
%(to be deleted?)}

From the start to the peak of the HXR bursts, which corresponds to the
time range from the dotted line to the thick gray line in Figure~5, both
the time profiles of the microwave spectral index $\delta_{\mu_{P}}$ and
that of the HXR spectral index $\delta_{X}$ decrease simultaneously,
keeping a certain gap $\Delta\delta$ of about 1.5.
%%%
The 2000 November 25 flare (event g) is a special case due to the very
long duration and the smooth variation of the HXR and microwave
emissions \citep{Taka07}.
This flare shows GH features even in the HXR spectral evolution, and is
a typical ``type-C flare'' (e.g. Hoyng et al. 1976), which have been
discussed from the {\it Hinotori} era
\citep{Cli86,Kai86,Dennis85,Kosu88}.
These imply that the magnetic trap effectively works both on the
microwave-emitting electrons and even on the HXR-emitting electrons.
The HXR-emitting electrons are trapped in the magnetic loop at least
once before they precipitate into the chromosphere.
The dominant microwave emission source is located at the footpoint
conjugated with the HXR source, which means that these
microwave-emitting electrons escape from the magnetic trap.
As \citet{Taka07} reported, on the other hand, the microwave footpoint
emission source disappears and the dominant emission source is from the
loop top during the valley times.
Therefore, the gaps $\Delta\delta$ increase during the valley times.

\subsection{Spatial Features of Spectral Index $\delta_{\mu}$}
%%%
The difference between the microwave emission sources and the HXR
emission sources implies that the microwave emissions mainly come from
electrons trapped magnetically within the flare loop that connects the
HXR footpoint sources.
Spatially resolved analyses on the spectral indices are, therefore,
required for these events, to explain the reason for the certain gap
$\Delta\delta$ of about 1.6 confirmed in current study.
By using the NoRH data, we examine the spatial distribution of the
spectral indices derived from microwaves, and compare the value at the
footpoint ($\delta_{\mu_{H}}$(X)) with that at the loop top
($\delta_{\mu_{H}}({\mu})$).

First, we derived the electron spectral index from NoRH microwave
emission $\delta_{\mu_{H}}$ at the brightest emission positions.
The brightest regions are determined to be regions where the intensities
at 17~GHz are larger than 80~\% of the maximum intensities, and the
derived spectral index is noted as $\delta_{\mu_{H}}({\mu})$.
%%%
For many cases in our results, the dominant microwave emissions are from
loop tops (or legs), and $\delta_{\mu_{H}}({\mu})$ show the values
there.
The twelfth column of Table~1 presents the list of
$\delta_{\mu_{H}}({\mu})$.
These items roughly correspond to those derived from the total intensity
$\delta_{\mu_{P}}$ (column 8) with the displacement of about 0.5.
The displacement is probably due to the spatial distribution of the
spectral index.
%%%
For the 2000 October 29 flare (event e), we failed to correctly derive
the $\alpha$-map, because of a large displacement between the position
of the NoRH 17~GHz emission source and that at NoRH 34~GHz.
The displacement is as large as 10$^{\prime\prime}$.
Although the reason is unknown, it may be caused by other energy release
processes that occurred in the preflare phase.
We omitted this event from the further discussions.

Second, we determined the spectral index at the HXR footpoints, that is,
$\delta_{\mu_{H}}$(X) from the NoRH $\alpha$-map.
If there are two HXR footpoint sources, we calculated them separately.
For events (f) and (h), we could not, however, correctly derive the
spectral index.
This is because the HXR footpoint positions are too close to the loop
top/leg microwave emission sources, and they are not spatially resolved.
For events (a) and (b), one of the two footpoints is too close to the
microwave emission source.
We showed the spectral indices $\delta_{\mu_{H}}$(X) for these footpoint
sources in parentheses in the eleventh column of Table~1.
The number I/II is the same as that marked for the HXR footpoints in the
Figure~1.
Here we have to note that the error of $\delta_{\mu_{H}}$(X).
The microwave emission from footpoints is weaker than that from loop
tops/legs for most cases, and therefore, the error to estimate the
spectral index $\delta_{\mu_{H}}$(X) could be larger than that for loop
top one ($\delta_{\mu_{H}}({\mu})$).
We roughly expect the error to be about $\pm$~0.7.
%%%
Except for the 2000 April 8 flare (event c), we found the relation that
the spectral indices for footpoint sources are larger, which corresponds
to the softer spectra, than those at the loop tops/legs, that is, 
$\delta_{\mu_{H}}({\rm X}) > \delta_{\mu_{H}}(\mu)$.
Especially, in events (d) and (g), $\delta_{\mu_{H}}({\rm X})$ is quite
close to the $\delta_{X}$.
%%%
We will discuss the spatial features of the spectral index
$\delta_{\mu}$ and the relation with the magnetic field in the next
section.

\section{Discussion and Conclusions}
%%%
We examined the electron spectral indices of nonthermal emissions seen
in HXRs ($\delta_{X}$) and in microwaves ($\delta_{\mu}$) for 12 flares
observed by {\it Yohkoh}/HXT NoRP, and NoRH.
%%%
Eight flares of the selected 12 events show gaps between the spectral
indices, i.e. $\Delta\delta$ of about 1.6.
The gaps are consistent with the result of \citet{Sil00}.
The other four events show larger gaps $\Delta\delta >$ 2.7, since they
suffer from softening of the HXR spectra (i.e. enlarging the HXR
spectral index $\delta_{X}$) due to the super-hot thermal component even
in the HXT M2 band.
In spite of the fact that we examined the spectral features for the
impulsive phase to avoid the effect of magnetic trapping, there still
remains a certain gap.
%%%
On the other hand, from the start to the peak of the HXR bursts (that
corresponds from the first vertical dotted line to the thick gray line
in Figure~5), both the time profile of the microwave spectral index
$\delta_{\mu_{P}}$ and that of the HXR spectral index $\delta_{X}$
decrease simultaneously, keeping a certain gap $\Delta\delta$ of about
1.6.
%%%
%The rest four flares, the gap $\Delta\delta$ is larger than 2.7.
%We conclude that this is because the HXR emissions suffered from the
%super-hot thermal components even in the M2 band, and the electron
%spectral index derived from HXRs $\delta_{X}$ softens.

We also investigated the positions of the emission sources by using the
HXT and NoRH data.
For five of the eight events, the brightest microwave emission sources
are located on the loop tops (or legs; group L).
On the other hand, for the other three events, they are different from
the HXR emission sources, which are mainly from footpoints (group
F/F$^{*}$)).
This implies that the microwave emissions mainly come from electrons
trapped magnetically within the flare loop that connects the HXR
footpoint sources.
%%%
The difference in the site of the emission sources possibly causes the
gap of the spectral indices $\Delta\delta$.
The spatial distribution of the microwave spectral index derived from
NoRH $\delta_{\mu_{H}}$ should therefore be examined.
%%%
Except for one event (event c), we confirmed that the microwave spectra
for footpoint sources are softer than those at the loop tops/legs, that
is, $\delta_{\mu_{H}}({\rm X}) > \delta_{\mu_{H}}(\mu)$.
However, the spatial distribution of the microwave spectral index again
cannot resolve the gap $\Delta\delta$ even at the sites of the HXR
emission sources.

From these results, we concluded that the spectra of the accelerated
electrons have a bent, and become harder above several hundreds of keV.
This is also consistent with results of previous studies
\citep{Yoshi85,Dennis88,Matsu05}.
%%%
\citet{Mino08} numerically calculated the spectral indices of microwave
and HXR emissions from the trapped and precipitating electrons, and
showed that a softer HXR spectrum and a hard microwave spectrum can be
generated.
The difference of the spectral indices $\Delta\delta$ $\sim$1.5, is
consistent with our result.
However, we showed that the spatially resolved distribution of the
microwave spectral index in our study cannot resolve the gap
$\Delta\delta$ even at the footpoint sources.
This means that there still remains a certain gap $\Delta\delta$.

Here, we examine events (c) and (d) in more detail.
Figure~6 shows the photospheric magnetograms for these events obtained
by the Michelson Doppler Imager (MDI; Scherrer et al. 1995) aboard the
{\it Solar and Heliospheric Observatory} ({\it SOHO}; Domingo et
al. 1995).
The levels of the contours are $\pm$400, $\pm$600, and $\pm$800 gauss
with the red and blue lines for positive and negative magnetic
polarities, respectively.
We overlaid the microwave contour image of NoRH 17~GHz on each panel
with {\it light blue} lines.
The HXR contour image observed with HXT in the M2 band is also overlaid
with {\it green} lines.
The levels of these contours for both contour images are 40, 60, 80, and
95~\% of the maximum intensities.
%%%
From Figure~6, we clearly see weak magnetic field strength at HXR
footpoints in event (c) of about 250 -- 300 gauss, while there is quite
a strong magnetic field in event (d) of about 900 -- 1000 gauss.
In event (d), therefore, we expect strong magnetic field for microwave
footpoint position, and it could be several 100 gauss.
The microwave gyrosynchrotron emission is strongly related to the
magnetic field strength, and only high-energy electrons can contribute
under the condition of weak magnetic field strength.
On the other hand, with strong magnetic field, weaker energy electrons
can contribute the emission.
As \citet{Bas99} calculated, electrons with the energy of about
500~keV contribute at 500 gauss, while the energy must be higher than
1~MeV at 200 gauss to generate 17~GHz emission.
This means that low energy electrons with the energy of about several
100~keV emit the gyrosynchrotron at the footpoints.
The energy of several 100~keV is as low as that for electrons emitting
HXR bremsstrahlung, which causes a very soft spectrum that has the same
spectral index $\delta_{\mu_{H}}$(X) as $\delta_{X}$.
The magnetic field strength at the loop top must be much smaller, and
the microwave-emitting electrons have high enough energy of about
several MeV, at which the harder spectral component appears.
%%%
On the other hand, in event (c), the magnetic field strength is weak
even at the HXR footpoint sources.
This means that high-energy electrons are responsible for the microwave
emission even at the footpoint, which corresponds to the observed harder
spectral component.
%%%
From these results, we concluded that the bent of the electron spectra
seems to be at about several 100~keV, which is consistent with the
previous suggestions \citep{Yoshi85,Dennis88}.
%%%
In Figure~3, we cannot see the clear bent of the HXR spectra derived
from WBS/HXS, and therefore, the bent of HXR spectra in the current case 
occurs at the energy higher than 300~keV.

\acknowledgments

We first would like to acknowledge an anonymous referee for her/his
comments and suggestions.
We would like to thank all the members of Nobeyama Solar Radio
Observatory, NAOJ for their supports during the observation.
We wish to thank Drs. S. Krucker and T. Minoshima for fruitful
discussions and his helpful comments.
This work was carried out by the joint research program of the
Solar-Terrestrial Environment Laboratory, Nagoya University.
The {\it Yohkoh} satellite is a Japanese national project, launched and 
operated by ISAS, and involving many domestic institutions, with 
multilateral international collaboration with the US and the UK.

{\it Facilities:} \facility{NoRH}, \facility{NoRP}, \facility{Yohkoh}.

\begin{deluxetable}{lccccccccc|cc}
\rotate
\tablecaption{List of Events with Their Spectral Characteristics.
\label{tbl-1}}
\tablewidth{0pt}
\tablehead{
\colhead{Data} & \colhead{Time$^{a}$} & \colhead{$\Delta t^{b}$} & 
\colhead{Position} & \colhead{GOES} & \colhead{Source$^{c}$} & 
\colhead{$\delta_{X}$} & \colhead{$\delta_{\mu_{P}}$} &
\colhead{$\Delta\delta^{d}$} & \colhead{Hardening$^{e}$} & 
\colhead{$\delta_{\mu_{H}}$(X)} & \colhead{$\delta_{\mu_{H}}(\mu)$}}
\startdata
(1) & (2) & (3) & (4) & (5) & (6) & (7) & (8) & (9) & (10) & (11) & (12) \\ \tableline 
(a)1998 Sep 09 & 04:56:31 & 5 & S17W70 & M2.8 & L & 5.0 & 2.8 & 2.2 & 0.45 & {\sc i }3.4 ({\sc ii }2.6) & 3.1 \\%& 7.9 \\
(b)2000 Jan 12 & 01:35:55*& 4 & N12E75 & M2.8 & L & 4.7 & 2.6 & 2.1 & 0.96 & {\sc i }2.7 ({\sc ii }1.9) & 2.2 \\%& 15.7 \\
(c)2000 Apr 08 & 02:38:37 & 3 & S15E28 & M2.0 & L & 5.2 & 3.1 & 2.1 & 1.27 & {\sc i }1.5, {\sc ii }1.5 & 2.1 \\%& 17.5 \\
%2000 Jul 25 & 02:47:48 & N04W05 & M8.0 & loop & 5.0 & 3.9 & 1.1 & --  \\
(d)2000 Sep 16 & 04:13:18 & 15 & N13W05 & M5.9 & L & 4.4 & 3.6 & 0.8 & --  & {\sc i }4.7, {\sc ii }3.8 & 2.8 \\%& 7.0 \\
(e)2000 Oct 29 & 01:47:06 & 5 & S21E34 & M4.4 & L?& 4.5 & 3.7 & 0.8 & 1.66 & -- & -- \\ %E3.0,W6.9 & 3.9 \\%& 11.0 \\ 
(f)2000 Mar 03 & 02:12:04*& 4 & S17W63 & M3.8 & F & 5.4 & 3.5 & 1.9 & 0.80 & (2.3) & 2.9 \\%& 15.1 \\
(g)2000 Nov 25 & 01:21:58 & 17 & N09E46 & M8.2 & F & 3.8 & 2.4 & 1.4 & 1.38 & 2.8 & 2.1 \\%& 5.2 \\
(h)2001 Mar 10 & 04:03:38 & 1 & N26W42 & M6.7 & F?& 3.9 & 2.2 & 1.7 & --  & (1.5) & 1.9 \\%& 15.9 \\ 
\tableline
(i)1998 Nov 28 & 05:40:27 & 21 & N21E41 & X3.3 & -- & 6.6 & 3.5 & 3.1 & -- & -- & -- \\
(j)1999 Aug 04 & 05:49:22 & 10 & S17W61 & M6.0 & -- & 6.2 & 2.9 & 3.3 & -- & -- & -- \\
(k)2001 Mar 30 & 05:13:15 & 3 & N18W18 & M2.2 & -- & 5.7 & 3.0 & 2.7 & -- & -- & -- \\
(l)2001 Sep 25 & 04:34:36 & 0 & S21E01 & M7.6 & -- & 6.5 & 3.6 & 2.9 & -- & -- & -- \\
\enddata
\tablenotetext{a}{\tiny Peak time of the HXR flux in the HXT M2 band in UT.
For events (b) and (f), we chose other preceding sub-peaks instead of
the maximum peaks.}
\tablenotetext{b}{\tiny Delay of the microwave peak times (17~GHz) to the
HXR (M2 band) peaks in second, $\Delta t = t_{\rm 17GHz} - t_{\rm M2}$.}
\tablenotetext{c}{\tiny Position of the microwave emission sources.
L and F mean loop top and footpoint sources, respectively.
For events (e) and (h), we cannot clearly classify.}
\tablenotetext{d}{\tiny Gap between the indices, $\Delta\delta = \delta_{X} -
 \delta_{\mu_{P}}$.}
\tablenotetext{e}{\tiny Hardening degree during the bursts.
See, text and Figure~6.}
\end{deluxetable}

\begin{figure}
\epsscale{0.8}
\plotone{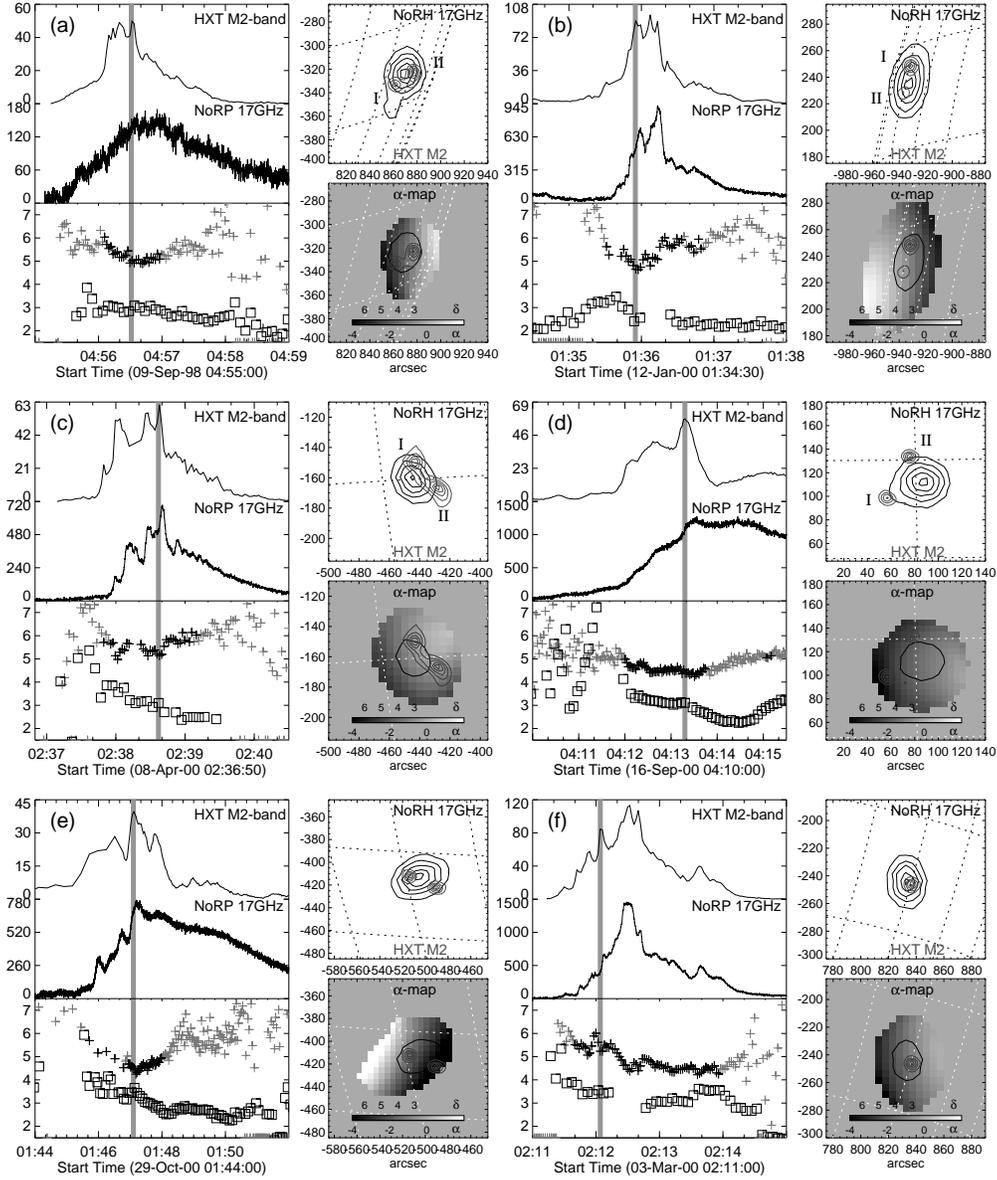}
\caption[1a]{(Figure 1a)
Time profiles and images for each selected event.
The intensity (top) and spectral (bottom) time profiles are shown on the
left-hand side of each sub-figure.
The HXR and microwave light curves are taken by the HXT M2 band (33 --
53 keV) and the NoRP 17~GHz, respectively.
The microwave and HXR spectral time profiles are plotted with the square
($\square$) and cross ($+$), respectively.
The 17~GHz contour images obtained by NoRH and the HXR ones obtained
with HXT (M2 band) are shown in the top right panel of each sub-figure
by the black and gray lines, respectively.
The contour levels of the microwave and HXR images are 20~\%, 40~\%,
60~\%, 80~\%, and 95\% of the peak intensity.
}
\end{figure}

\begin{figure}
\epsscale{0.8}
\plotone{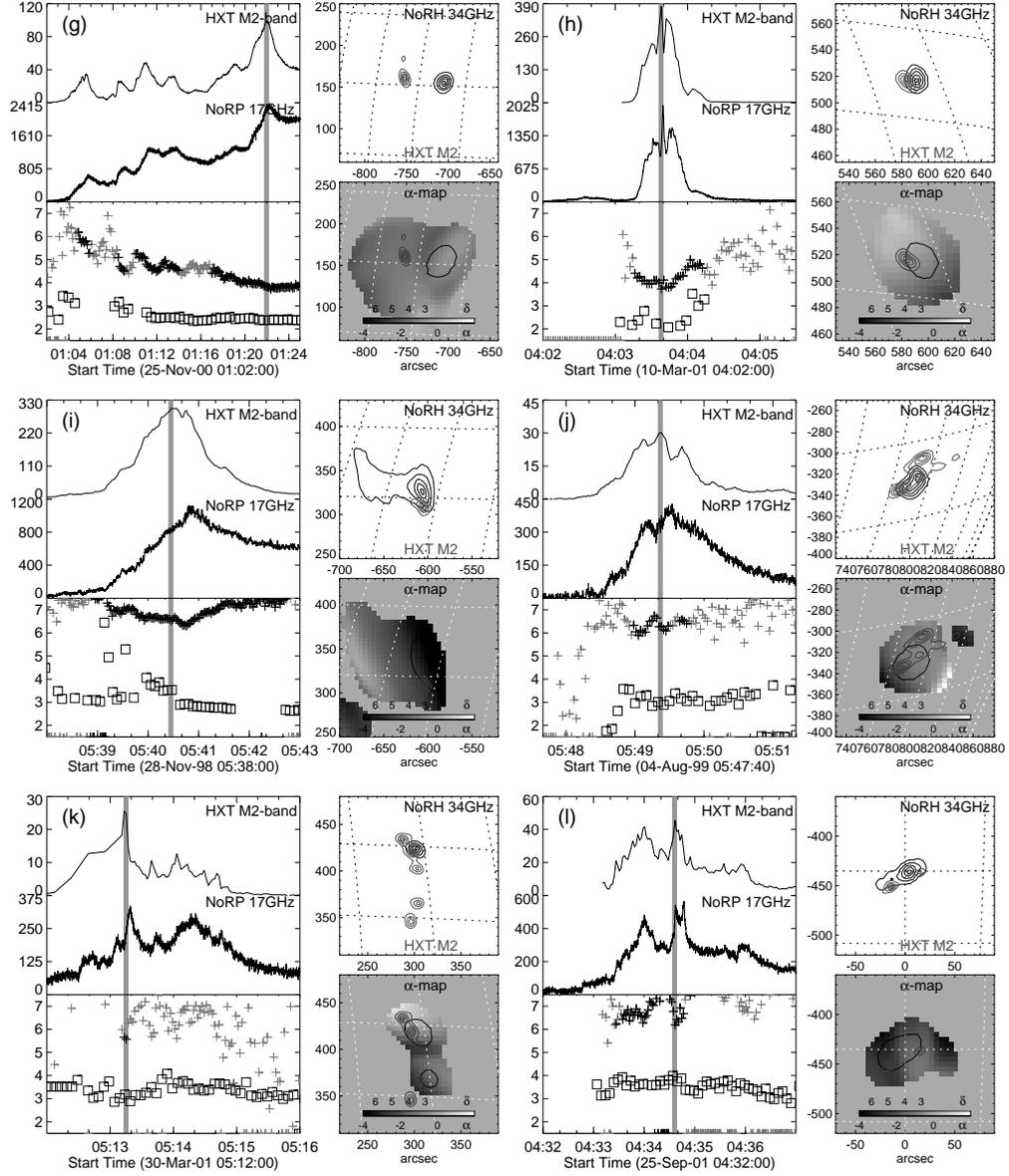}
\caption[1b]{(Figure 1b) -- continued.}
\end{figure}

\begin{figure}
\epsscale{1.0}
\plotone{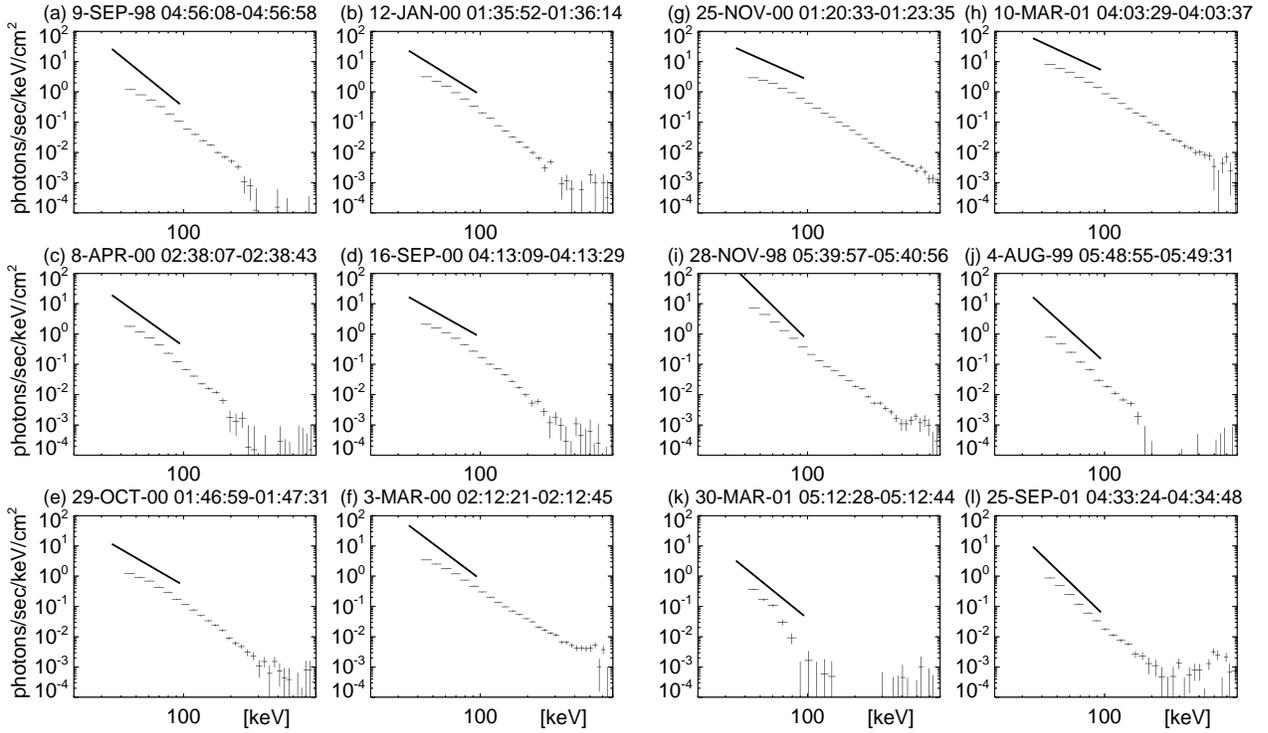}
\caption[2]{(Figure 2)
HXR spectra derived from WBS/HXS.
The times on the top are the integration times in UT.
The slopes of the power-law distributions derived from HXT are overlaid
with the solid lines.
}
\end{figure}

\begin{figure}
\epsscale{0.8}
\plotone{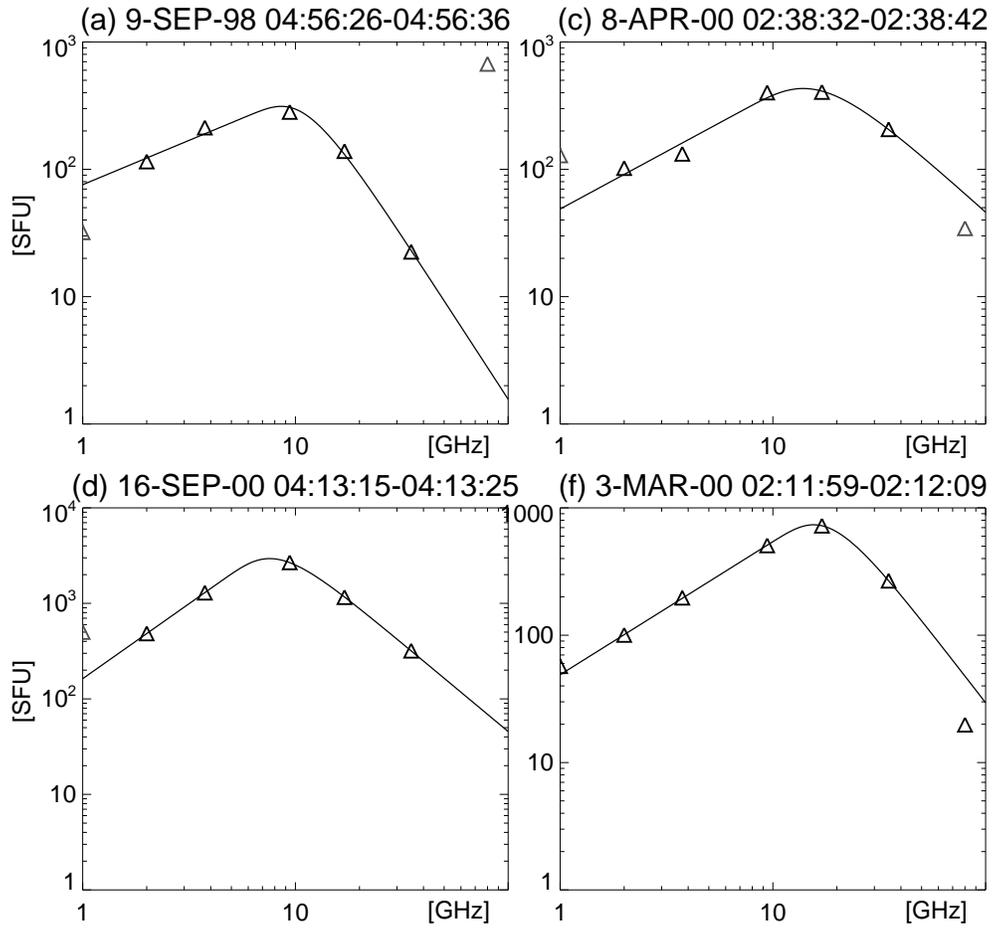}
\caption[3]{(Figure 3)
The microwave spectra taken by NoRP for events (a), (c), (d), and (f).
The solid lines are the fitting results.
}
\end{figure}

\begin{figure}
\epsscale{0.8}
\plotone{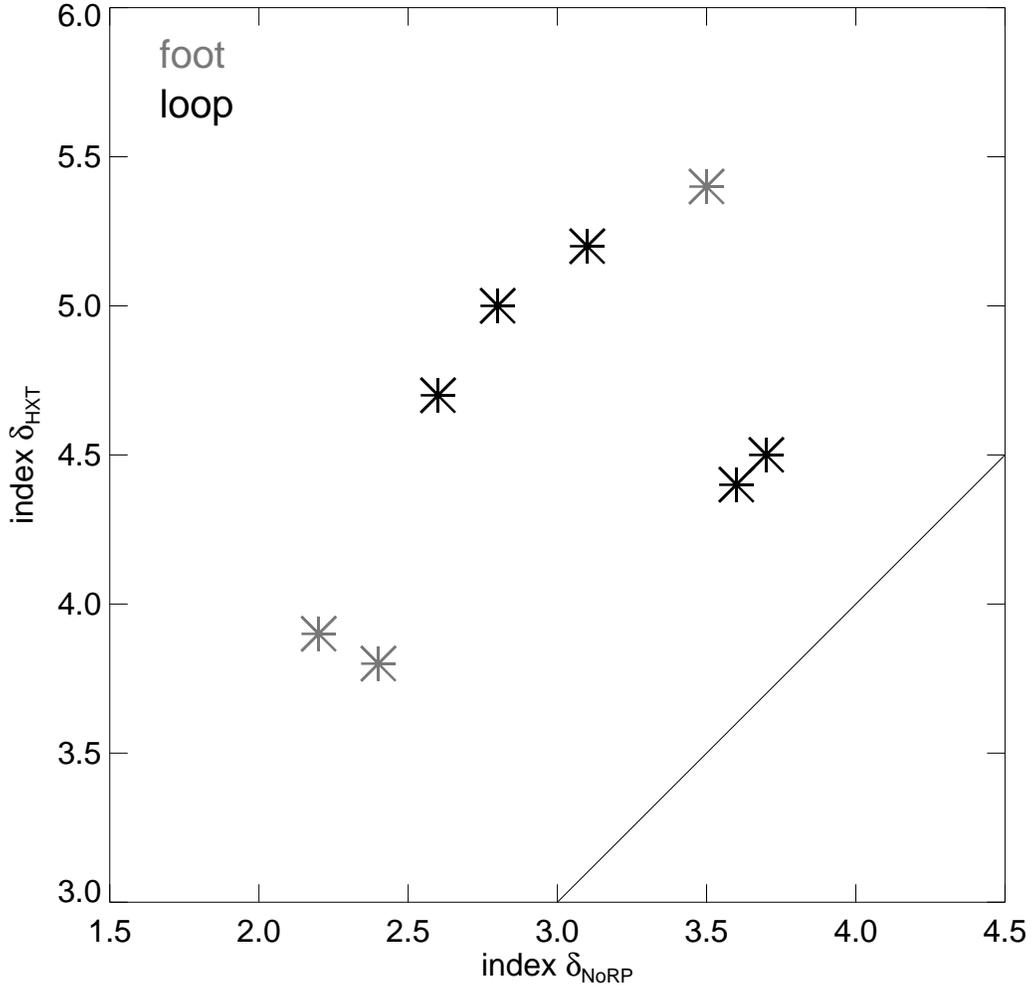}
\caption[4]{(Figure 4)
Scatter plot between the electron spectral index derived from microwaves
$\delta_{\mu}$ (horizontal axis) and that from HXRs $\delta_{X}$
(vertical axis).
The solid line shows the points where $\delta_X$ corresponds to
$\delta_{\mu}$ ($\Delta\delta = 0$).
Only the small gap events ($\Delta\delta < 2.2$, see text) are plotted.
}
\end{figure}

\begin{figure}
\epsscale{1.0}
\plotone{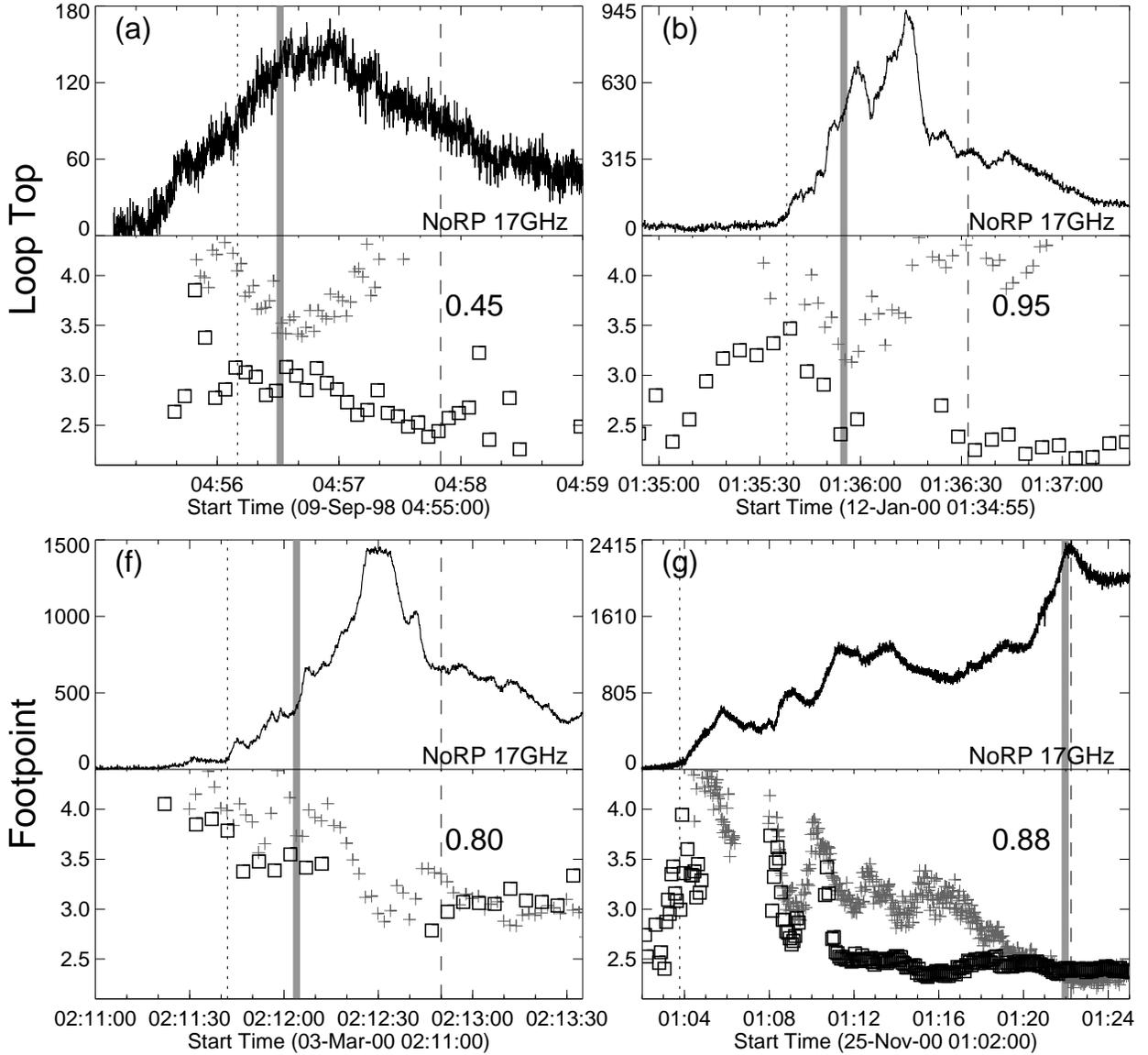}
\caption[5]{(Figure 5)
Hardening of microwave spectral index $\delta_{\mu_{P}}$ for events (a),
(b), (f), and (g).
The vertical dotted lines correspond to the times when the microwave
17~GHz fluxes taken by NoRP exceed 100~SFU (Solar Flux Unit).
The vertical dashed lines correspond to the times when the 17~GHz fluxes
become half of the peak values, except for event (g), which show
 saturating the microwave spectral index.
The bottom left panels show the time profiles of the spectral index
$\delta_{\mu_{P}}$ with the square ($\square$) marks.
The overlaid time profiles are the spectral index $\delta_{X}$ (cross; $+$).
Note, we subtracted 1.5 from the original value of the index to clearly
show the temporal variation.
The numbers noted in the figure show how much the spectral indices
$\delta_{\mu_{P}}$ decrease during the two vertical dotted lines.
}
\end{figure}

\begin{figure}
\epsscale{1.0}
\plotone{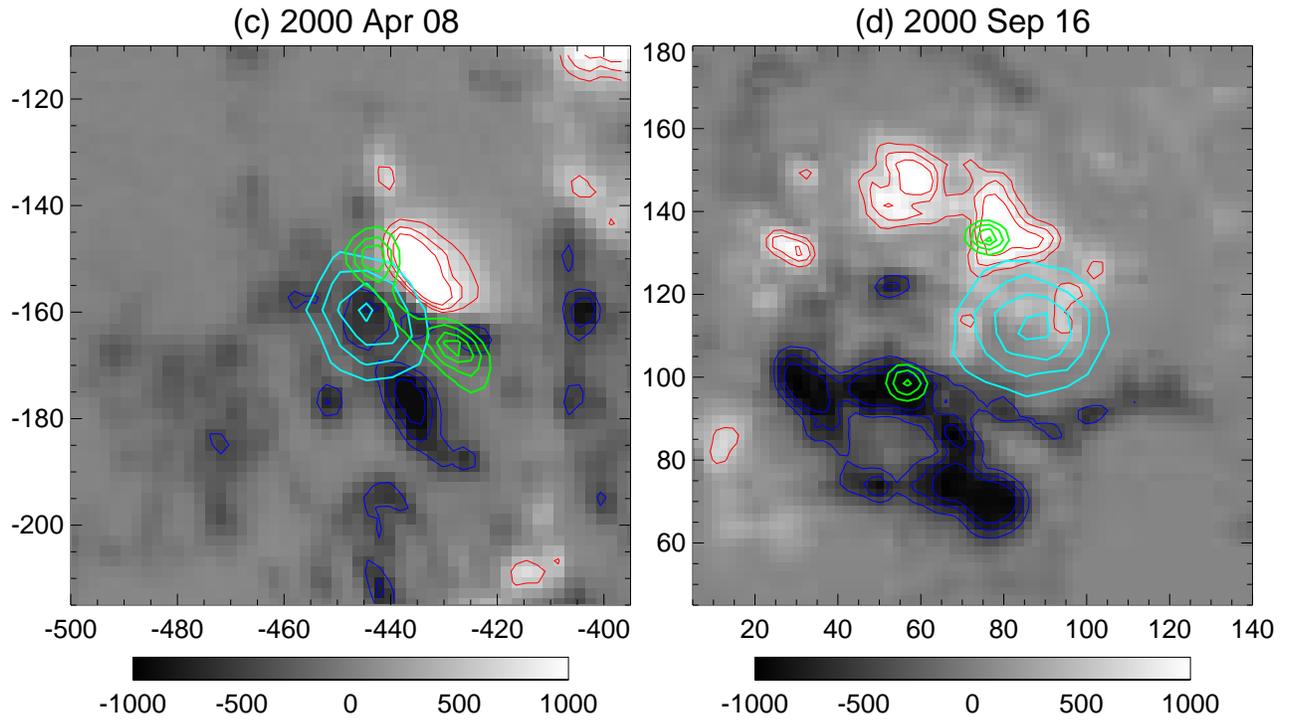}
\caption[6]{(Figure 6)
Magnetic field strengths for events (c) and (d).
The levels of the contours are 400, 600, and 800 gauss with the red and
blue lines for positive and negative magnetic polarities, respectively.
The microwave contour image of NoRH 17~GHz on each panel is overlaid
with the {\it light blue} lines.
The HXR contour image observed with HXT in the M2 band is also overlaid
with the {\it green} lines.
The levels of these contours for both images are 40, 60, 80, and 95~\%
of the maximum intensities.
}
\end{figure}

\end{document}